# Reaching the unreached

## A Role of ICT in sustainable Rural development.

| Mr.Nayak S.K. | Dr.S.B.Thorat | Dr.Kalyankar N.V. |
|---|---|---|
| Head, Dept. of Computer Science | Director, | Principal |
| Bahirji Smarak Mahavidyalaya, | Institute of Technology and Mgmt. | Yeshwant Mahavidyalaya, Nanded |
| Basmathnagar, Dist.Hingoli. (MS), | Nanded, Dist.Nanded. (MS), | Nanded (MS) |
| India | India | India |

*Abstract—*

We have seen in last few decades that the progress of information technology with leaps and bounds, which have completely changed the way of life in the developed nations. While internet has changed the established working practice and opened new vistas and provided a platform to connect, this gives the opportunity for collaborative work space that goes beyond the global boundary.

ICT promises a fundamental change in all aspects of our lives, including knowledge dissemination, social interaction, economic and business practices, political engagement, media, education, health, leisure and entertainment.

In India ICT applications such as Warana, Dristee, Sari, Sks, E-Chaupal, Cybermohalla, Bhoomi, E-Mitra, Deesha, Star, Setu, Friends, E-Seva, Lokmitra, E-Post, Gramdoot, Dyandoot, Tarahaat, Dhan, Akshaya, Honeybee, Praja are in functioning for rural development.

ICT offers an opportunity to introduce new activities, new services and applications into rural areas or to enhance existing services. With more than 70% of the Indian population living in rural areas and earns its live hood by agriculture and allied means of income.

ICTs can play a significant role in combating rural and urban poverty and fostering sustainable development through creating information rich societies and supporting livelihoods. If ICTs are appropriately deployed and realize the differential needs of urban and rural people, they can become powerful tools of economic, social and political empowerment.

This paper introduces the application of ICT for rural development. The paper aims at improving the delivery of information to rural masses such as: technology information, marketing information, and information advice. This paper focuses digital divide and poverty eradication, good governance and the significance of internet for rural development. The paper concludes that ICTs offer the developing country, the opportunity to look ahead several stages of rural development by the use of internet. Effective use of ICT can demolish geographical boundaries and can bring rural communities closer to global economic systems and be of meaningful help to the underprivileged.

## I. INTRODUCTION

### A. ICT Defined

ICT can be interpreted broadly as "technologies that facilitate communication and the processing and transmission of information by electronic means."

### B. ICT and Development

According to Nelson Mandela, "Eliminating the distinction between information rich and information poor countries is critical to eliminating other inequalities between North and South and to improving the quality of life of all humanity." (Mandela, 1998).

### C. Rural India

"Just as the whole universe is contained in the Self, so is India contained in the villages"…This has been said by **Mahatma Gandhi**, the father of our nation and the visionary architect of India's rural development programme. The villages epitomize the soul of India. Rural India reflects the very essence of Indian culture and tradition.

## II. INTERNET

Internet is another great achievement of ICT. Internet is one of the technologies available for global resources and information haring. Internet has become a common resource of the whole mankind, inquiring and sharing information become easier than ever. With Internet, geographical distance and state borders are eliminated.

The Internet can be a powerful democratizing force, offering greater economic, political and social participation to communities that have traditionally been undeserved and helping developing nations meet pressing needs. It needs to ensure that everyone has a chance to share in the benefits of the Digital Age, information technology.

Internet can be a tool for rural development.

TABLE I.     WORLD-WIDE INTERNET USERS PENETRATION

| Country | Users (In Millions) | Population (In Millions) | Percentage (As on 07/12/09) |
|---|---|---|---|
| India | 81.00 | 1156.89 | 7.00 % |
| USA | 227.71 | 307.21 | 74.10 % |
| Japan | 95.97 | 127.07 | 75.52 % |
| China | 360.00 | 1338.61 | 26.30 % |
| UK | 46.68 | 61.11 | 76.40 % |
| Germany | 54.22 | 82.32 | 65.90 % |
| Canada | 25.08 | 33.48 | 9.80 % |
| France | 43.10 | 62.15 | 69.30 % |
| Australia | 17.03 | 21.26 | 61.00 % |
| Russia | 45.25 | 140.04 | 32.31 % |
| Indonesia | 30.00 | 240.27 | 12.50 % |

www.internetworldstats.com/stats.htm

Above table depicts that there is need to promote internet uses in developing country like India.

## III.     IMPACT OF ICT ON SOCIETY

History has seen a move from the agricultural society, through the industrialized society and to the information society. Society is better informed as a result of developments in ICT. Now a day ICT is used in daily practices. For ex. training, education and entertainment these help society to a great extent. There is impact of increased availability of information using ICT on a variety of public services.

### A.   Digital Divide

ICTs threaten to expand the already wide socio-economic gap between urban and rural populations in developing countries. Simultaneously offering opportunities to reduce it. ICT implementation can be successfully adapted for the development of challenges faced by rural communities. Such implementations can be evaluated so that the often unexpected and desirable results that emerge can be revealed and accounted for.

### B.   E-Governance and Empowerment

The poverty can be adequately addressed by effective use of e-governance and ICT application in environmental management. Improved governance by using ICT can have direct impact in reducing poverty and improving the environment.

ICT can contribute fostering empowerment and participation by making government processes more efficient and transparent by encouraging communication and information sharing among rural and marginalized people.

### C.   ICT and Agriculture

The vast majority of poor people lives in rural areas and derives their livelihoods directly or indirectly from agriculture. ICTs can deliver useful information to farmers about agriculture like crop care and animal husbandry, fertilizer and feedstock inputs, pest control, seed sourcing and market prices.

The ICT application in agriculture sector is still very limited in India, and full range of potential benefits that such technology can provide us yet to be realized.

### D.   ICT and Poverty Alleviation

ICT applications in developing countries are often part of an overall strategy for economic growth. The role of ICT in poverty reduction is not limited to reducing income poverty, but also includes non-economic dimensions - in particular empowerment.

### E.   ICT and Health

Health care is one of the most promising areas for poverty alleviation. ICTs are being used in India to facilitate remote consultation, diagnosis and treatment.

Delivering health care with ICTs enables health care professionals and institutions to address the critical medical needs of rural communities, especially those in remote locations and those that lack qualified medical personnel and services.

### F.   ICT for Education

Moreover, appropriate use of ICTs in the classroom fosters critical, integrative and contextual teaching and learning; develops information literacy (the ability to locate, evaluate and use information). Thus, it improves the overall efficiency of the delivery of education in schools and educational management institutions at the national, state/provincial and community level.

The use of ICTs in education aims to improve the quality of teaching and learning as well as democratize the access to education.

### G.   ICT for Economic Development

Information and Communication Technology has a vital role in connecting the rural community to outside world for exchange of information, a basic necessity for economic development. Effective use of ICT can demolish geographical boundaries and can bring rural communities closer to global economic systems and be of meaningful help to the underprivileged.

### H.   Employment Opportunities

Poor people in rural localities have lack of opportunities for employment because they often do not have access to information about them. One use of ICTs is to provide on-line services for job placement through electronic labor exchanges in public employment service or other placement agencies.

## IV.     CHALLENGES AND ISSUES

The barriers that currently limit the development of rural areas include Distance barriers: to access to administrative and governmental structures,     Economic barriers: to access to wider business and labor markets, Social barriers: to information, education facilities, health and social services,

Information barriers : many rural areas and their amenities are undiscovered, unknown for the outer world.

For successful implementation of E- rural development it is necessary to define ICT requirements and issues for rural areas like vision, policy, awareness, technology, Infrastructure, services, applications. It is extremely important to focus on all aspects of rural development. For introduction and implementation of E-rural policy it is necessary to establish cooperation of different sectors of government and on national level to introduce coherent E-rural policy, establish formal platform, supporting exchange information and knowledge which will join researchers, developers, regional and local government and which could coordinate research and implementation activity and which can also support E-rural policy

Information-and-ICT initiatives are political. The effectiveness and potential of ICTD initiatives can be inhibited or circumscribed by national and/or local power relations. Political awareness and analysis is an important aspect of ICT4D planning at all levels.

Harnessing ICTs for human development requires awareness-raising and constituency-building across all levels of society.

The challenge for governments is to ensure the convergence of their initiatives and those taken up by various donors, multilaterals, NGOs and other organizations and to address the digital divide.

## V.   DISCUSSION / CONCLUSIONS

### A.   Agriculture

- The farmers are also not able to know about the prices prevailing in other markets, as the Market Committees are able to disseminate information mostly in respect of their own markets.

- The availability of prompt and reliable market information about what is happening in the market, what quantities are arriving and what prices are quoted for different commodities considerably improves the decision making capability of the farmers and strengthens their bargaining power.

### B.   Education

- With introduction of ICT based education at school level our children and youngsters will grow as "Computer kids". Their exposure will get increased due to which the knowledge level will get definitely improved.

- It enhances the quality of education.

### C.   Health Services

- Delivering health care with ICTs enables health care professionals and institutions to address the critical medical needs of rural communities, especially those in remote locations and those that lack qualified medical personnel and services.

- General knowledge and awareness regarding health can be promoted among the people.

### D.   Commerce and Trade

- Still the benefits are not percolating down to the farmers, as they are unable to plan their strategies for sale of their produce at remunerative prices, in the absence of correct and timely market information and advice about arrivals, prices, market trend, etc.

- In terms of market opportunities emerging agricultural technologies are   increasingly information intensive and the rural poor must now cope with increasingly sophisticated input and output markets.

### E.   Local Governance and Community life

- ICT can empower rural communities (particularly marginalized group) and give them "a voice" which permits them to contribute to the development process.

- With ICT government can serve the public better. Computers and the Internet make it possible for government to contact and transact with lots of people at the same time.

## VI.   SUGGESTIONS / RECOMMENDATIONS

- ICT should be harnessed for the benefit of ordinary farmers.

- ICT policy is the most important factor on the introduction of information communication technology to the rural mass with a view to empower them by providing all types of information and communication with regard to their day to day life.

- Awareness building on ICT usage should be conducted to the rural mass through the institutions such as gram panchayat, government office, schools, ICT Centers etc.

- Government has to encourage the software developers to develop software packages in their respective mother tongue.

- While the dedicated and qualified English teachers are appointed to the schools to enhance the English knowledge. Computer knowledge has to be given in their mother tongue with the support of English upto the students are able to understand and handle the computers.

- Established cyber cafes in rural areas to facilitate the rural people who cannot purchase computers by themselves. To encourage the private sector to establish cyber café to provide ICT services in the rural areas.

- It is necessary to open dialogue and professional discussions to create awareness on ICT in the rural areas.

- To support and facilitate to develop locally relevant contents for the Internet.

- Significant attention needs to pay for the development of business models for content creation in the rural computing context.

## VII.  STRATEGIES

- Adopt one village one computer scheme.

- Start Community learning and Information Centers (CLIC) centers.

- Establish Market Information Centers in remote areas.

- Establish Tele Centers in remote areas.

- Encourage the use of computers and Internet in rural areas.

- Establish Information Technology (IT) parks in remote areas.

- Government has to reduce taxation of ICT-related components, products and services.

- Establish partnerships with NGOs engaged in awareness and innovative for ICT4RD.

- Explore the use of Free and Open Source Software (FOSS).

- Explore the use of local language software's.

- Promote the benefits of ICTs to private sector and academic institutions, and encourage computerization.

- Begin basic ICT skills workshops for all rural students at tertiary level.

- Encourage ICT awareness programmes, especially among primary and secondary school students in rural areas.

- Promote ICT-related courses at university/college level and expand the base of supportive certificate and diploma level at college level.

- Connect schools, universities and research centers to national and international distance education facilities, national and international databases, libraries, research laboratories and computing facilities.

- Encourage corporations to appreciate ICT competent staff and conduct/sponsor ICT training for staff members/professionals.

- Encourage assessment and promotion of civil servants to include ICT competency.

- Encourage ICT4D research and development and partnership with the private sector and international educational/research centers.

## VIII.  THE ROAD AHEAD

Rural Development forms an important agenda of the Government. However, the uptake of e-governance in the Rural Development sector has been relatively slow. The main reasons for this are poor ICT infrastructure in rural areas, poor ICT awareness among agency officials working in rural areas and local language issues. Efforts are, however, on to extend infrastructure up to village level. Already, many states have gone ahead to provide connectivity up to block level. This has helped in taking the e-governance efforts further closer to the people.

The important requirement of establishing infrastructure in rural areas is now being taken up as a high-agenda project after the President of India envisioned the idea of providing urban amenities in rural areas (PURA). PURA (Provision of urban amenities in Rural Areas) has been conceived as a scheme under MoRD and envisages to achieve its objective by bridging the various kinds of divide that exists between rural and urban areas by providing four major kinds of connectivity to rural areas: physical (road, power), electronic (telecommunication, internet), knowledge and market. With the provision of such connectivity, it is hoped that the benefits of e-governance in the Rural Development sector would reach its true beneficiaries.

Crucial success factors to realize this dream are strong political & administrative will, Government Process Reform, capacity building of provider (government functionaries) and consumer (rural citizens), utilization of ICTs as a medium to share information and deliver services that are demand-driven and people-centric.

## ACKNOWLEDGMENT (*HEADING 5*)

We are thankful to Hon. Ashok Chavan (Chief Minister, Maharashtra) India, Society members of Shri. Sharada Bhawan Education Society, Nanded. Also thankful to Shri. Jaiprakash Dandegaonkar (Ex-State Minister, Maharashtra), Society members of Bahiri Smarak Vidyalya Education Society, Wapti for encouraging our work and giving us support.

Also thankful to our family members and our students.

AUTHORS PROFILE

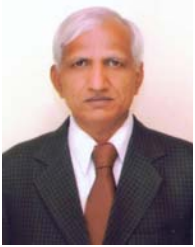

**Dr. N.V.Kalyankar**
Principal
Yeshwant Mahavidyalaya, Nanded (Maharashtra)

Completed M.Sc. (Physics) from Dr.B.A.M.U, Aurangabad. In 1980 he joined as a lecturer in department of physics at Yeshwant Mahavidyalaya, Nanded. In 1984 he completed his DHE. He completed his Ph.D. from Dr.B.A.M.U, Aurangabad in 1995. From 2003 he is working as a Principal to till date in Yeshwant Mahavidyalaya, Nanded. He is also research guide for Physics and Computer Science in S.R.T.M.U, Nanded. He is also worked on various bodies in S.R.T.M.U, Nanded. He also published research papers in various international / national journals. He is peer team member of NAAC (National Assessment and Accreditation Council, India). He published a book entitled "DBMS concepts and programming in FoxPro". He also got "Best Principal" award from S.R.T.M.U, Nanded in 2009. He is life member of Indian National Congress, Kolkata (India). He is also honored with "Fellowship of Linnean Society of London (F.L.S.)" on 11 November 2009.

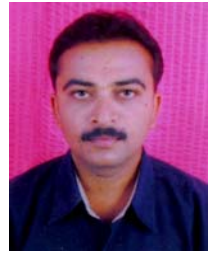

**S.K.Nayak**
M.Sc. (Computer Science), D.B.M, B.Ed.

He completed M.Sc. (Computer Science) from S.R.T.M.U, Nanded. In 2000 he joined as lecturer in Computer Science at Bahirji Smarak Mahavidyalaya, Basmathnagar. From 2002 he is acting as a Head of Computer Science department. He is doing Ph.D. He attended many national and international conferences, workshops and seminars. He is having 2 international publications. His interested areas are ICT, Rural development, Bioinformatics.

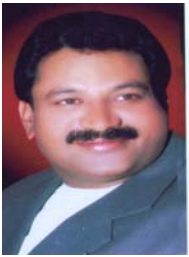

**Dr.S.B.THorat**
M.E. (Computer Science & Engg.)
M.Sc. (ECN), AMIE, LM-ISTE, Ph.D. (Comp.Sci. & Engg.)

He is having 24 years teaching experience. From 2001 he is working as a Director, at ITM. He is Dean of faculty of Computer studies at Swami Ramanand Teerth Marathwada University, Nanded (Maharashtra). Recently he is completed his Ph.D. He attended many national and International conferences. He is having 7 international publications. His interested area are AI, Neural network, Data mining, Fuzzy systems, Image processing.